\documentclass[twocolumn,a4paper,reprint,amssymb,aps,prd,groupedaddress,superscriptaddress,nofootinbib]{revtex4-2}
\pdfoutput=1
\usepackage{booktabs}
\usepackage{multirow}
\usepackage{physics}
\usepackage{dcolumn}
\usepackage{bm}
\usepackage{amsmath}
\usepackage{mathtools}
\usepackage{amsfonts}
\usepackage{pbox}
\usepackage{color}
\usepackage[dvipsnames]{xcolor}
\usepackage{tabularx}
\usepackage{subfigure}
\usepackage{graphicx}
\usepackage{psfrag}
\usepackage{lipsum}
\usepackage{enumitem}
\usepackage{changes}
\usepackage{array}
\usepackage{CJK}
\usepackage[english]{babel}

\usepackage[colorlinks = true,
            linkcolor = blue,
            urlcolor  = blue,
            citecolor = blue,
            anchorcolor = blue]{hyperref}
\usepackage{blindtext}
\usepackage[utf8]{inputenc}
\usepackage{graphicx}
\usepackage{placeins}
\usepackage{siunitx}
\usepackage{amssymb}
\usepackage{mathtools}
\usepackage{physics}
\usepackage{tikz}
\usepackage[titletoc]{appendix}
\usepackage{listings}
\usepackage{float}
\usepackage{braket}

\usepackage[capitalize]{cleveref}
\usepackage{soul}
\usepackage{array}

\begin{document}
\title{Testing spatial curvature in an anisotropic extension  of $w$CDM model with low redshift data}
\author{Vikrant Yadav}
\email{vikuyd@gmail.com}
\affiliation{School of Basic and Applied Sciences, Raffles University, Neemrana - 301705, Rajasthan, India}

\author{Rajpal}
\email{rajpal05041985@gmail.com}
\affiliation{School of Basic and Applied Sciences, Raffles University, Neemrana - 301705, Rajasthan, India}

\author{Pardeep}
\email{pardeep.math.rs@igu.ac.in}
\affiliation{Department of Mathematics, Indira Gandhi University, Meerpur, Haryana 122502, India}

\author{Manish Yadav}
\email{manish.math.rs@igu.ac.in}
\affiliation{Department of Mathematics, Indira Gandhi University, Meerpur, Haryana 122502, India}

\author{Santosh Kumar Yadav}
\email{sky91bbaulko@gmail.com}
\affiliation{School of CS \& AI, SR University, Warangal-506371, Telangana, India}

\begin{abstract}

In this letter, we report the observational constraints on a Bianchi type I anisotropic extension of $w$CDM model with spatial curvature from observational data including Baryon Acoustic Oscillations (BAO), Cosmic chronometers (CC), Big Bang nucleosynthesis (BBN), Pantheon+ (PP) compilation of SNe Ia and SH0ES Cepheid host distance anchors. The anisotropy is found to be of the order $10^{-13}$, which interplay with spatial curvature to reduce $H_0$ tension by $\sim 1\sigma$ as found in the analyses with BAO+CC+BBN+PP combination of data, while no significant effect of anisotropy is observed with BAO+CC+BBN+PPSH0ES combination of data. A closed Universe is favored by $w$CDM as well as anisotropic $w$CDM models with spatial curvature in analyses with BAO+CC+BBN+PP combination of data. An observation of an open Universe from $w$CDM model with spatial curvature in analyses with BAO+CC+BBN+PPSH0ES combination of data and a closed Universe from anisotropic $w$CDM model with curvature in analyses with same combination of data is made. The quintessence form of dark energy is favored at 95\% CL in both analyses.

\end{abstract}
\maketitle
\section{Introduction}
\label{sec:intro5}

One of the basic geometric quantities of the Universe is spatial curvature which we shall denote by $\kappa$. As spatial flatness $\kappa = 0$ is predicted by the most widely accepted models of early Universe inflation, any finding of nonzero spatial curvature, henceforth referred to as curvature, would have a profound effect on our comprehension of the early Universe and cosmic development. Furthermore, the validity of the Robertson-Walker (RW) metric and the cosmic homogeneity and isotropy may be questioned if measurements of $\kappa$ from observations at various redshifts (or in other directions) greatly diverged from one another. Numerous techniques have been put forth for measuring and assessing curvature. The model independent method proposed by \cite{Denissenya2018CosmicCT} determines curvature at a level of $6 \times 10^{-3}$ by means of strong lensing time delays and supernova distance measurements. The inclusion of curvature in cosmological tests of General Relativity (GR) is emphasised by \cite{Dossett2012SpatialCA} because it can lead to apparent departures from GR and reduce restrictions on changed growth parameters. The difficulties of detecting curvature with high accuracy are discussed in \cite{Leonard:2016evk}, where it is suggested that such measurements may not be possible without making significant assumptions about Dark Energy (DE) development and the standard $\Lambda$CDM (Cosmological Constant $\Lambda$ + Cold Dark Matter) parameter values. A test particle approach for measuring the curvature is proposed by \cite{Ciufolini1986HowTM}; it requires a minimum of four particles in vacuum and a minimum of six particles in general. Numerous research have suggested model-independent techniques that use Cosmic Chronometers (CC) and gravitational-wave standard sirens to determine cosmic curvature. The potential of these approaches is seen in \cite{Wei:2018cov} and \cite{He:2021rzc}, where \cite{He:2021rzc} indicates that a dependable limit on cosmic curvature can be obtained from the DECIGO space-based observatory. The applications of CC and gravitational waves are further explored in \cite{Zheng:2019trp} and \cite{Wang:2022rvf}, whereby \cite{Wang:2022rvf} demonstrates that cosmic curvature may be measured using gravitational waves and the distance sum rule in strong gravitational-lensing. Model-independent techniques for estimating cosmic curvature are presented by \cite{Wei:2019uss, Wang:2020dbt}, and also, \cite{Wang:2020dbt} employing machine learning to do so. When paired with CC and gravitational-wave standard sirens, these investigations show that model-independent techniques have great promise for detecting cosmic curvature. Furthermore, recent research has investigated the possibility of expanding the $\Lambda$CDM model to incorporate anisotropic expansion and non-zero curvature.  Shear-free (SF) metrics, with particular signals on the Cosmic Microwave Background (CMB) temperature spectrum, were suggested by \cite{Pereira2016ExtendingT} as a feasible substitute to explain the large-scale development of the cosmos. A stress on the transition redshift was reported by \cite{Arjona:2021hmg}, who also proposed new consistency tests for curvature and homogeneity. A closed $\Lambda$CDM model with non-zero curvature might somewhat ease some cosmic tensions, according to research done on the Planck 2015 data \cite{Ooba2018Planck2C}. According to \cite{Akarsu:2021max}, a minor degree of expansion anisotropy cannot be ruled out even if the observational data still indicate spatial flatness and isotropic expansion. All these studies point to the possibility of investigating non-zero curvature and anisotropic expansion further, even though the $\Lambda$CDM model is still viable. For some recent studies regarding curvature and/or anisotropy, see \cite{DiValentino:2019qzk,Yang:2022kho,Handley:2019tkm,DiValentino:2020srs,Virey:2008nu,DiDio:2016ykq,Mortsell:2011yk,Akarsu:2019pwn,Akarsu:2021max,Amirhashchi:2018bic,Amirhashchi:2018nxl,Amirhashchi:2020qep,Abdalla:2022yfr,DiValentino:2020vhf,Yadav:2023yyb} and references therein.\\ 

In this letter, following our previous work \cite{Yadav:2024pvr}, we report the observational constraints on a Bianchi type I anisotropic extension of the $w$CDM model with curvature from low redshift observational data as described in the following sections. Our main purpose is to assess the impact/interplay of spatial curvature and anisotropy in the dynamics of the $w$CDM Universe. 
\section{Model, datasets and methodology}
\label{sec:2}
In continuation of our recent work \cite{Yadav:2024pvr}, we consider the $w$CDM model with curvature and denote this model as $w$CDM+$\Omega_{\rm \kappa0}$, where $\Omega_{\rm \kappa0}$ represents the curvature parameter, and it is the ratio of the present-day actual curvature density of the Universe to the critical density, which is required for a flat Universe. Further, we consider the anisotropic extension of $w$CDM+$\Omega_{\rm \kappa0}$, that is, $w$CDM+$\Omega_{\rm \kappa0}$+$\Omega_{\rm \sigma0}$, where $\Omega_{\rm \sigma0}$ is the expansion anisotropy parameter. The governing Friedmann equation reads \cite{Akarsu:2021max,Yadav:2024pvr}

\begin{eqnarray}\label{model} 
\frac{H^2}{H_0^2}&  =&\Omega_{\sigma0}a^{-6}+\Omega_{\text{r}0}a^{-4}+\Omega_{\rm de0}a^{-3(1+w_{\rm de0})}\\\nonumber
&&+\Omega_{\kappa0}a^{-2}.   
\end{eqnarray} 
 Here $a$ is the average scale factor; $H$ is the average Hubble parameter with $H_0$ being its present-day value, describing the expansion rate of the Universe. Further,   $\Omega_{\rm r0}$, $\Omega_{\rm m0}$, $\Omega_{\kappa0}$, and $\Omega_{\rm de0}$ denote the radiation, matter, curvature, and dark energy density parameters, respectively. These parameters satisfy the equation 
 $$\Omega_{\sigma0} + \Omega_{\rm r0}+ \Omega_{\rm m0}+  \Omega_{\rm de0} + \Omega_{\kappa0} = 1.$$

In the following, very briefly, we describe the employed data sets and the chosen methodology for carrying out the observational analysis of the model under consideration. For more details, we refer the reader to our previous work \cite{Yadav:2024pvr}. \\

\noindent\textbf{Baryon Acoustic Oscillation (BAO)}: We use 14 BAO measurements from the completed Sloan Digital Sky Survey (SDSS)~\cite{eBOSS:2020yzd}. These include the independent BAO measurements of angular-diameter distances and Hubble distances relative to the sound horizon from eight different samples, using the galaxies, quasars, and Lyman-$\alpha$ (Ly$\alpha$) for completed experiments. We use the sound horizon's comoving size ($r_{\rm s}$) at the drag redshift ($z_{\rm d}$), given by 
\begin{equation}{
\label{comoving size}
r_{\rm d}=r_{\rm s}(z_{\rm d})=\int_{z_{\rm d}}^\infty \frac{c_{\rm s}\text{d}z}{H(z)}.}
\end{equation}
Here $c_{\rm s}=\frac{c}{\sqrt{3(1+\mathcal{R})}}$ is the baryon-photon fluid's sound speed. Further, $\mathcal{R}=\frac{3\Omega_{\rm b0}}{4\Omega_{\rm \gamma 0}(1+z)}$, where $\Omega_{\rm b0}=0.022h^{-2}$ and $\Omega_{\gamma 0}=2.469\times 10^{-5}h^{-2}$ are the present-day physical densities of baryons and photons, respectively,  with $h=H_0/100$ being the reduced Hubble constant ~\cite{Cooke:2016rky,Bennett:2020zkv}. Following \cite{Yadav:2023yyb,Yadav:2024pvr}, we allow $z_d$ to vary in our analyses with BAO data as opposed to the related previous studies of \cite{Akarsu:2019pwn,Akarsu:2021max}.
\\

\noindent\textbf{Cosmic Chronometer (CC)}: 
We use CC data of 33 $H(z)$ measurements spanning over the redshift values from 0.07 to 1.965 \cite{Zhang:2012mp,Simon:2004tf,Moresco:2012jh,Ratsimbazafy:2017vga,Stern:2009ep,Borghi:2021rft,Jiao:2022aep,Moresco:2015cya}, which provide a basic relationship between cosmic time $t$, redshift $z$, and the Hubble parameter $H(z)$ \cite{Jimenez:2001gg}: $H(z)= -\frac{1}{1+z}\frac{\text{d}z}{\text{d}t}$.\\

\noindent\textbf{Big Bang Nucleosynthesis (BBN)}: 
We employ an updated BBN estimate of the physical baryon density, $\omega_b$ (where $\omega_b \equiv \Omega_bh^2$) from experimental nuclear physics at the Laboratory for Underground Nuclear Astrophysics (LUNA) of the INFN Laboratori Nazionali del Gran Sasso in Italy ~\cite{Mossa:2020gjc}, with a value of $0.02233\pm0.00036$. \\

\noindent\textbf{Pantheon+ and SH0ES}:
 We use the distance moduli measurements obtained from the supernovae of Type Ia (SNe Ia).  The theoretical apparent magnitude $m_B$ of a supernova at redshift $z$ reads:
\begin{eqnarray}
\label{distance_modulus}
m_B = 5 \log_{10} \left[ \frac{d_L(z)}{1\rm Mpc} \right] + 25 + M_B,
\end{eqnarray}
where $M_B$ is the absolute magnitude, and $d_L(z)$ is the luminosity distance.

We utilize SNe Ia distance modulus data from the Pantheon+ sample \cite{Brout:2022vx} with 1701 light curves corresponding to 1550 distinct supernovae Ia in the redshift range of $z \in [0.001, 2.26]$, and refer to this collection as PP. When we incorporate the SH0ES Cepheid host distance anchors \cite{Brout:2022vx} into our analyses, we refer this dataset to as PPSH0ES.\\

%%%%%%%%%%%%%%%%%%%%%%%%%%%%%%%%%%%%%%%%%%%%%%%%%%%%%%%%%%%%%%%%%%%%%%%%%%%%%%%%%%%%%%%%%%%%%%%%%%%%%%%%%%%%%%%%
The model's set of free baseline parameters is provided by 
$$\mathcal{P}_{w\rm CDM+\Omega_{\kappa0}+\Omega_{\rm \sigma0}}= \left\{ \omega_{\rm b}, \, \omega_{\rm c}, \, H_0,  \, w_{\rm de0}, \Omega_{\kappa0}, \, \Omega_{\sigma0 }  \right\}.$$
Here $\omega_{\rm b}=\Omega_{\rm b} h^2$ and $\omega_{\rm c}=\Omega_{\rm c}h^2$ are  physical density parameters of baryons and cold dark matter, respectively, in the present Universe.

  We use uniform priors: $\omega_{\rm b}\in[0.01,0.03]$, $\omega_{\rm c}\in[0.05,0.25]$, $\,H_0\in[60,80]$, $w_{\rm de0} \in [-2,0]$, $\Omega_{\kappa0 }\in[-0.3,0.3]$, $\Omega_{\sigma0 }\in[0,0.001]$. Employing the aforementioned datasets, viz., BAO, CC, BBN and PP\&SH0ES, We obtain the correlated Monte Carlo Markov Chain (MCMC) samples from the interface of \texttt{MontePython}~\cite{Audren:2012wb} with the publicly available Boltzmann code \texttt{CLASS}~\cite{Blas:2011rf}. Further, the MCMC samples are further analyzed using the python package \texttt{GetDist}\footnote{\href{https://getdist.readthedocs.io/en/latest/intro.html}{https://getdist.readthedocs.io/en/latest/intro.html}}.

\section{Results and Discussion}
\label{sec4}
%%%%%%%%%%%%%%%%%%%%%%%%%%%%%%%%%%%%%%%%%%%%%%%%%%%%%%%%%%%%%%%%%%%%%%%%%%%%%%%%%%%%%%%%%%%%%%%%%%%%%%%%%%%%%%%%%%%%%%%%%%%%%%%%%%%%%%%%%%%%%%%%%%%%%%%%%%%%%%%%%%%%%%%%%%%%%%%%%%%%%%%%%%%%%%%%%%%%%%%%%%%%%%%%%%%%%%%%%%%%%%%%%%%%%%%%%%%%%%%%%%%%%%%%%%%%%%%%%%%%%%%%%%%%%%%%%%%%%%%%%%%%%%%%%%%%%%%%%%%%%%%%%%%%%%%%%%%%%%%%%%%%%%%%%%%%%%%%%%%%%%%%%%%%%%%%%%%%%%%%%%%%%%%%%%%%%%%%%%%%%%%%%%%%%%%%%%%%%%%%%%%%%%%%%%%%%%%%%%%%%%%%%%%%%%%%%%%%%%%%%%%%%%%%%%%%%%%%%%%%%%%%%%%%%%%%%%%%%%%%%%%%%%%%%%%%%%%%%%%%%%%%%%%%%%%%%%%%%%%%

The observational constraints on free parameters and some derived parameters of the $w$CDM+$\Omega_{\rm \kappa0}$ and $w$CDM+$\Omega_{\rm \kappa0}$+$\Omega_{\rm \sigma0}$ models at $68\%$ CL from BAO+CC+BBN+PP and BAO+CC+BBN+PPSH0ES combinations of data are presented in Table \ref{tab:1_5}. The constraints (at $68\%$CL) on the $\Lambda$CDM+$\Omega_{\rm \kappa0}$ and $\Lambda$CDM+$\Omega_{\rm \kappa0}$+$\Omega_{\rm \sigma0}$ models from the same set of  combinations of data are represented by the second row above every parameter (highlighted in blue) for the purpose of comparison. The upper bounds of $\Omega_{\rm \sigma0}$ at 95\% CL are of order $10^{-13}$ for the $w$CDM+$\Omega_{\rm \kappa0}$+$\Omega_{\rm \sigma0}$ model with both combinations of data. The upper bound of $\,\,$ $\Omega_{\rm \sigma0}$ (at $\,\,$ 95\% CL) for this model with $\,\,$ BAO+CC+BBN+PP (BAO+CC+BBN+PPSH0ES)  $\,\,$ combination of data reads as $\,\,$ $\Omega_{\rm \sigma0} < 3.50\times 10^{-13}$ ($\Omega_{\rm \sigma0}<3.90\times 10^{-13}$). Similar upper bounds of $\Omega_{\rm \sigma0}$, at 95\% CL, are observed for $\Lambda$CDM+$\Omega_{\rm \kappa0}$+$\Omega_{\rm \sigma0}$ model which read as $\Omega_{\rm \sigma0} < 3.60\times 10^{-13}$ and $\Omega_{\rm \sigma0} < 3.90\times 10^{-13}$ with BAO+CC+BBN+PP and BAO+CC+BBN+PPSH0ES combinations of data, respectively.  

%%%%%%%%%%%%%%%%%%%%%%%%%%%%%%%%%%%%%%%%%%%%%%%%%%%%%%%%%%%%%%%%%%%%%%%%%%%%%%%%%%%%%%%%%%%%%%%%%%%%%%%%%%%%%%%%%%%%%%%%%%%%%%%%
\begin{table*}[ht]
  \caption{Constraints (mean values with 68\% CL errors) on  the free and some derived $\,\,$ parameters of the $w$CDM+$\Omega_{\rm \kappa0}$ and $w$CDM+$\Omega_{\rm \kappa0}$+$\Omega_{\rm \sigma0}$ models at 68\% CL from $\,\,$ BAO+CC+BBN+PP and BAO+CC+BBN+PPSH0ES combinations of data. The upper bounds on $\Omega_{\rm \sigma0}$ are displayed at 95\% CL.  The $\,\,$ Hubble constant $H_0$ is measured in the unit of $\,\,$ $\rm km\, s^{-1} \rm Mpc^{-1}$. The constraints on $\Lambda$CDM+$\Omega_{\rm \kappa0}$ and $\Lambda$CDM+$\Omega_{\rm \kappa0}$+$\Omega_{\rm \sigma0}$ (highlighted in blue colour) model parameters are also displayed.}
  \label{tab:1_5}
  \scalebox{1.1}{
	\begin{centering}
	  \begin{tabular}{lcccc}
  	\hline
  	 \hline
    \multicolumn{1}{l}{Data set} & \multicolumn{2}{c}{BAO+CC+BBN+PP} & \multicolumn{2}{c}{BAO+CC+BBN+PPSH0ES} \\  \hline
    &$w$CDM+$\Omega_{\rm \kappa0}$ & $w$CDM +$\Omega_{\rm \kappa0}$ +$\Omega_{\sigma0}$& $w$CDM+$\Omega_{\rm \kappa0}$ & $w$CDM+$\Omega_{\rm \kappa0}$ +$\Omega_{\sigma0}$   \\
    
    %& \textcolor{red}{$w$CDM }& \textcolor{red}{$w$CDM+$\Omega_{\sigma0}$} & \textcolor{red}{$w$CDM}  & \textcolor{red}{$w$CDM+$\Omega_{\sigma0}$}    \\
    
    & \textcolor{blue}{$\Lambda$CDM+$\Omega_{\rm \kappa0}$} & \textcolor{blue}{$\Lambda$CDM+$\Omega_{\rm \kappa0}$+$\Omega_{\sigma0}$} & \textcolor{blue}{$\Lambda$CDM+$\Omega_{\rm \kappa0}$} & \textcolor{blue}{$\Lambda$CDM+$\Omega_{\rm \kappa0}$+$\Omega_{\sigma0}$}    \\
\hline
%\vspace{0.1cm}    
      
 $10^{-2}\omega_{\rm b}$ & $2.238^{+0.035}_{-0.035}$ &  $2.230^{+0.035}_{-0.035}$ & $2.240^{+0.036}_{-0.036}$ &   $2.230^{+0.036}_{-0.036}$ \\

    & \textcolor{blue}{$2.238^{+0.036}_{-0.036}$} & \textcolor{blue}{$2.228^{+0.034}_{-0.039}$} & \textcolor{blue}{$2.241^{+0.034}_{-0.034}$} &  \textcolor{blue}{$2.232^{+0.036}_{-0.036}$} \\

  \hline
  \vspace{0.1cm}

  $\omega_{\rm c }$  &$0.264^{+0.034}_{-0.035}  $ & $0.221^{+0.044}_{-0.045} $ & $0.298^{+0.025}_{-0.026} $ & $0.238^{+0.041}_{-0.043}$
  \\
  & \textcolor{blue}{$0.261^{+0.033}_{-0.034}$} & \textcolor{blue}{$0.217^{+0.041}_{-0.043} $} & \textcolor{blue}{$0.295^{+0.025}_{-0.024}$} & \textcolor{blue}{$0.236^{+0.043}_{-0.044}$}
  \\
 \hline
 \vspace{0.10cm} 
  
 $\Omega_{\rm \sigma0 }$ & 0 & $< 3.50\times 10^{-13}$ & 0 & $<3.90\times 10^{-13}$    \\

   &   \textcolor{blue}{0} & \textcolor{blue}{$<3.60\times 10^{-13}$ } & \textcolor{blue}{0} &\textcolor{blue}{ $<3.90\times 10^{-13}$}  
   \\

   \hline
 \vspace{0.1cm} 
 
$\Omega{}_{\kappa0}$ & $-0.069^{+0.081}_{-0.081}$ & $0.032^{+0.086}_{-0.098}$  & $-0.188^{+0.066}_{-0.066}$ & $-0.042^{+0.081}_{-0.081}$    \\

   &   \textcolor{blue}{$0.026^{+0.044}_{-0.044}$} & \textcolor{blue}{$0.141^{+0.054}_{-0.063}$ } & \textcolor{blue}{$-0.084^{+0.029}_{-0.029}$} &\textcolor{blue}{$0.069^{+0.061}_{-0.061}$}  
   \\

\hline
 \vspace{0.1cm}

 $w_{\rm de0}$  &   $-0.892^{+0.076}_{-0.052}$  &  $-0.871^{+0.084}_{-0.058}$  & $-0.892^{+0.063}_{-0.046}$ &   $-0.860^{+0.066}_{-0.053}$\\

 &   \textcolor{blue}{-1} & \textcolor{blue}{-1} & \textcolor{blue}{-1} &\textcolor{blue}{-1}  
   \\

 \hline

 \vspace{0.10cm}

$H_{\rm 0}$  &    $67.2^{+1.6}_{-1.6}  $  & $68.2^{+1.7}_{-1.7}  $    & $ 72.19^{+0.87}_{-0.87} $ & $ 72.25^{+0.84}_{-0.84} $\\

   & \textcolor{blue}{$67.3^{+1.6}_{-1.6}$} & \textcolor{blue}{$68.3^{+1.5}_{-1.8} $ }  & \textcolor{blue}{$ 72.27^{+0.85}_{-0.85} $} & \textcolor{blue}{$72.48^{+0.89}_{-0.89} $}\\

\hline
 \vspace{0.1cm}
 
$M_{B}$  & $-19.436^{+0.048}_{-0.048}  $ & $-19.398^{+0.051}_{-0.051} $ & $-19.292^{+0.024}_{-0.024}$ &           $-19.281^{+0.023}_{-0.023}$ \\

 &  \textcolor{blue}{$-19.439^{+0.049}_{-0.049} $} & \textcolor{blue}{ $-19.403^{+0.050}_{-0.050} $ }&\textcolor{blue}{$-19.296^{+0.023}_{-0.023}$} & \textcolor{blue}{$-19.284^{+0.024}_{-0.024}$} \\

\hline
 \vspace{0.10cm} 

 $\Omega_{\rm m0}$ &   $0.265^{+0.017}_{-0.017}   $   & $0.224^{+0.023}_{-0.023}  $  & $0.299^{+0.013}_{-0.013} $ & $0.239^{+0.022}_{-0.022} $     \\

   &  \textcolor{blue}{$0.312^{+0.015}_{-0.015}$}   & \textcolor{blue}{$0.266^{+0.023}_{-0.019} $} & \textcolor{blue}{$0.340^{+0.012}_{-0.012}$} & \textcolor{blue}{$0.280^{+0.022}_{-0.022}$}    \\

\hline
 \vspace{0.10cm}

 $z_{\rm d}$ &   $1059.8^{+1.2}_{-1.2}$   & $1058.5^{+1.3}_{-1.3}$ & $1062.47^{+0.91}_{-0.91} $ & $1060.0^{+1.2}_{-1.2}$     \\

    &  \textcolor{blue}{ $1059.8^{+1.2}_{-1.2} $ }  & \textcolor{blue}{$1058.2^{+1.2}_{-1.2} $ } & \textcolor{blue}{$1062.43^{+0.89}_{-0.89}$} & \textcolor{blue}{$1060.1^{+1.2}_{-1.2}$}    \\
    
 \hline
 \vspace{0.10cm}

 $r_{\rm d}$ &   $147.4^{+3.3}_{-3.3}$   & $144.6^{+3.3}_{-3.3} $ & $138.7^{+1.9}_{-1.9} $ &  $137.5^{+1.9}_{-1.9}  $   \\

    &  \textcolor{blue}{$147.6^{+3.3}_{-3.3}$}  & \textcolor{blue}{$145.0^{+3.3}_{-3.3}$} &  \textcolor{blue}{$138.9^{+1.8}_{-1.8}$} & \textcolor{blue}{$137.8^{+1.9}_{-1.9} $} \\

    \hline
 \vspace{0.10cm}

 $\chi^2_{\rm min}$ &   $1435.4$   &  $1434.22$  & $1322.94$ & $1319.06$ \\

     &  \textcolor{blue}{$1438.28$}   & \textcolor{blue}{$1437.26$}  & \textcolor{blue}{$1326.56$} & \textcolor{blue}{$1322.78$} \\
\hline
\hline

\end{tabular}
\end{centering}}

\end{table*}

%%%%%%%%%%%%%%%%%%%%%%%%%%%%%%%%%%%%%%%%%%%%%%%%%%%%%%%%%%%%%%%%%%%%%%%%%%%%%%%%%%%%%%%%%%%%%%%%%%%%%%%%%%%%%%%%%%%%%%%%%%%%%%%%%%

\begin{figure*}[hbt!]
	\centering
	\includegraphics[width=5.5cm]{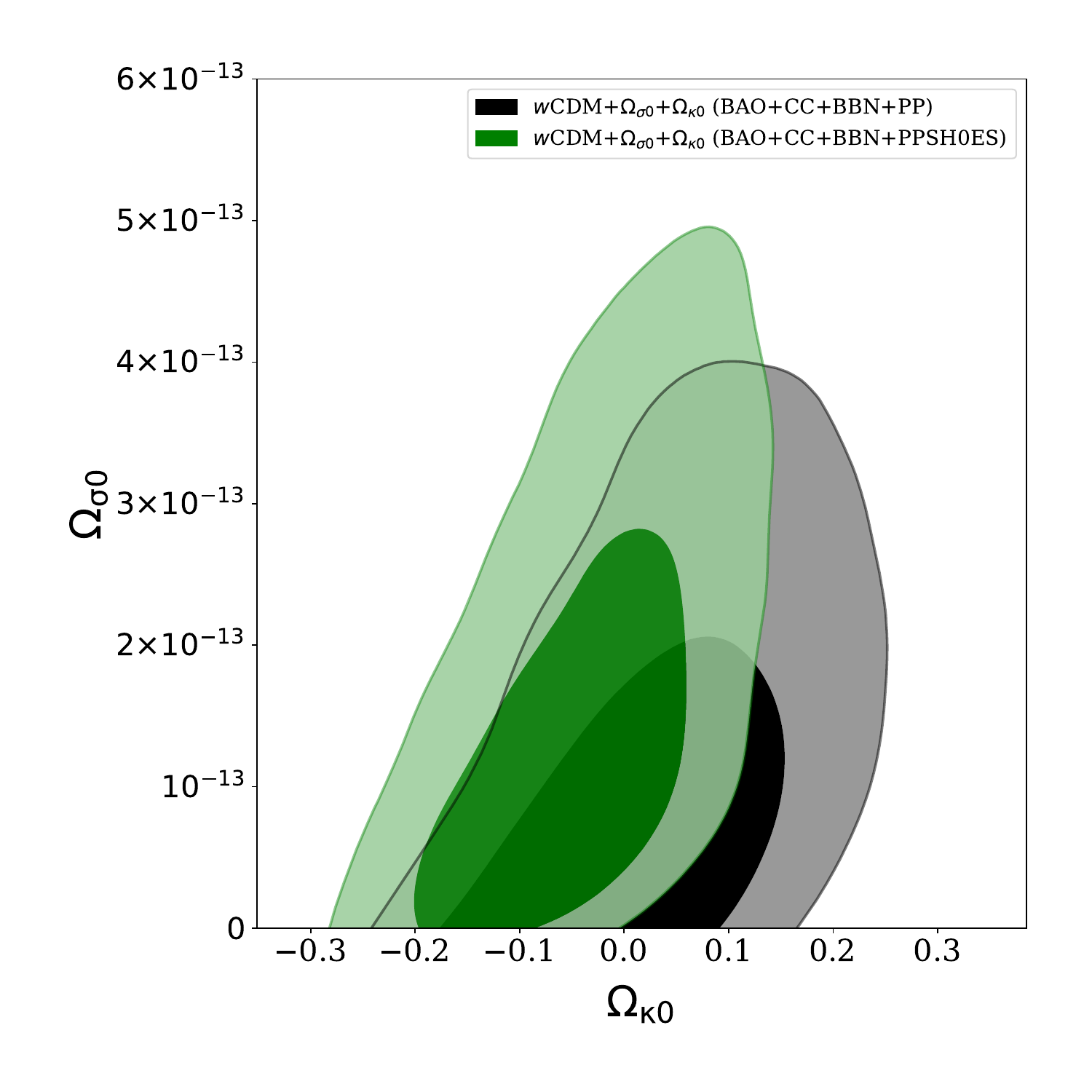}
 	\includegraphics[width=5.5cm]{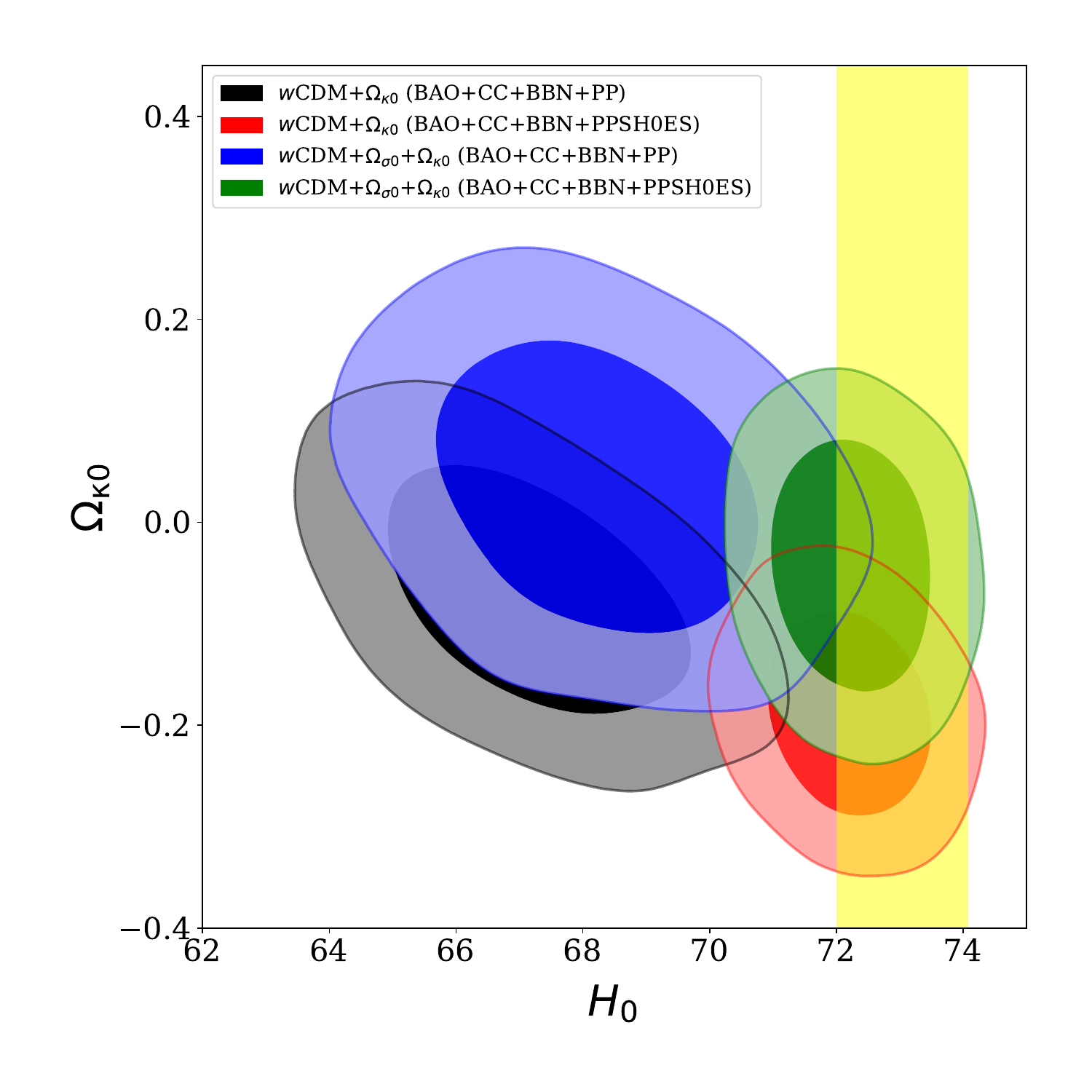}
  	\includegraphics[width=5.5cm]{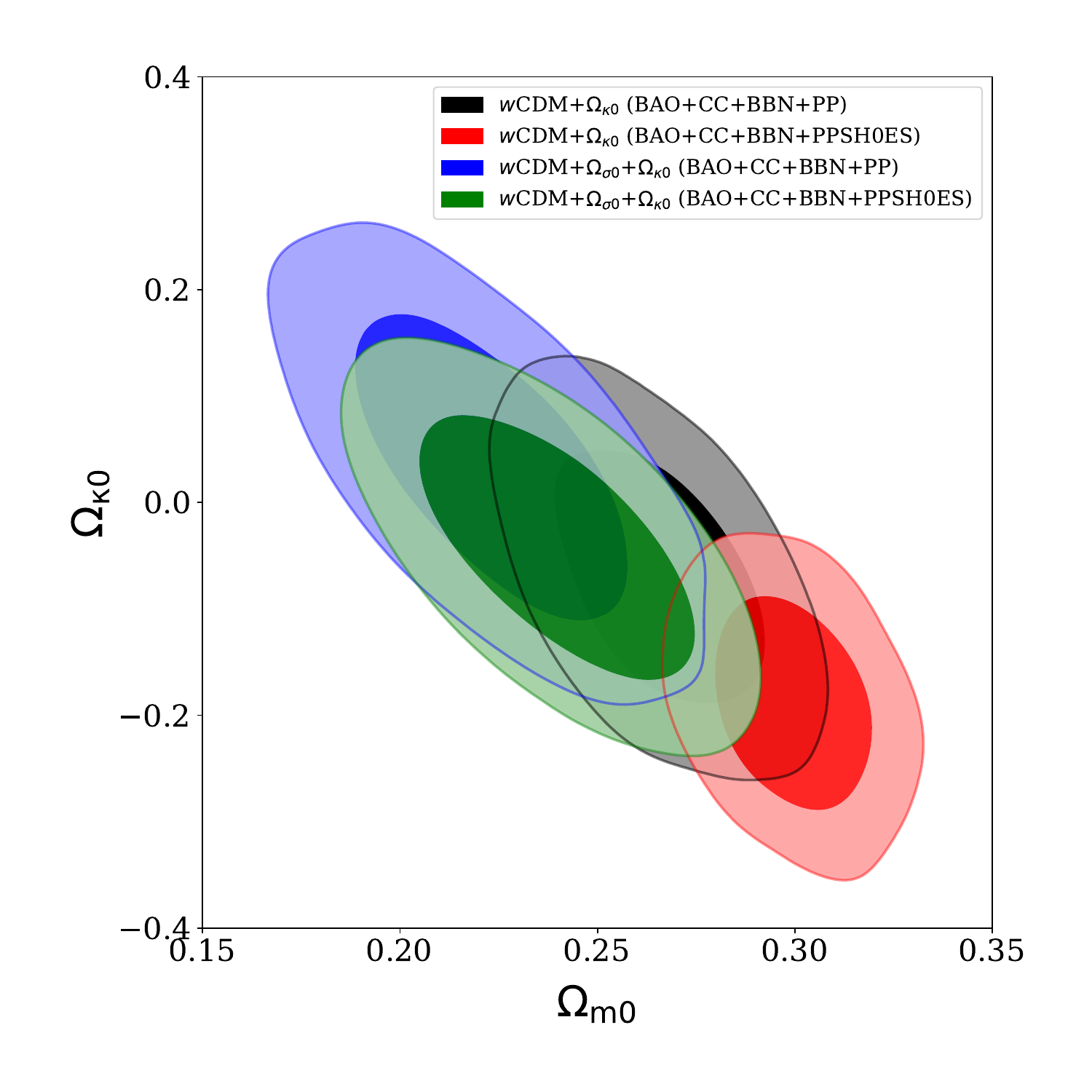}
	\caption{Two-dimensional marginalized confidence regions (at $\,\,$ 68\% and 95\% CL) $\,\,$ of some parameters of interest pertaining to $w$CDM+$\Omega_{\rm \kappa0}$ and $w$CDM+$\Omega_{\rm \kappa0}$+$\Omega_{\rm \sigma0}$  models $\,\,$ from $\,\,$ BAO+CC+BBN+PP and BAO+CC+BBN+PPSH0ES data. } 
	\label{fig2_5}
\end{figure*}

\begin{figure*}[hbt!]
	\centering
	\includegraphics[width=16cm]{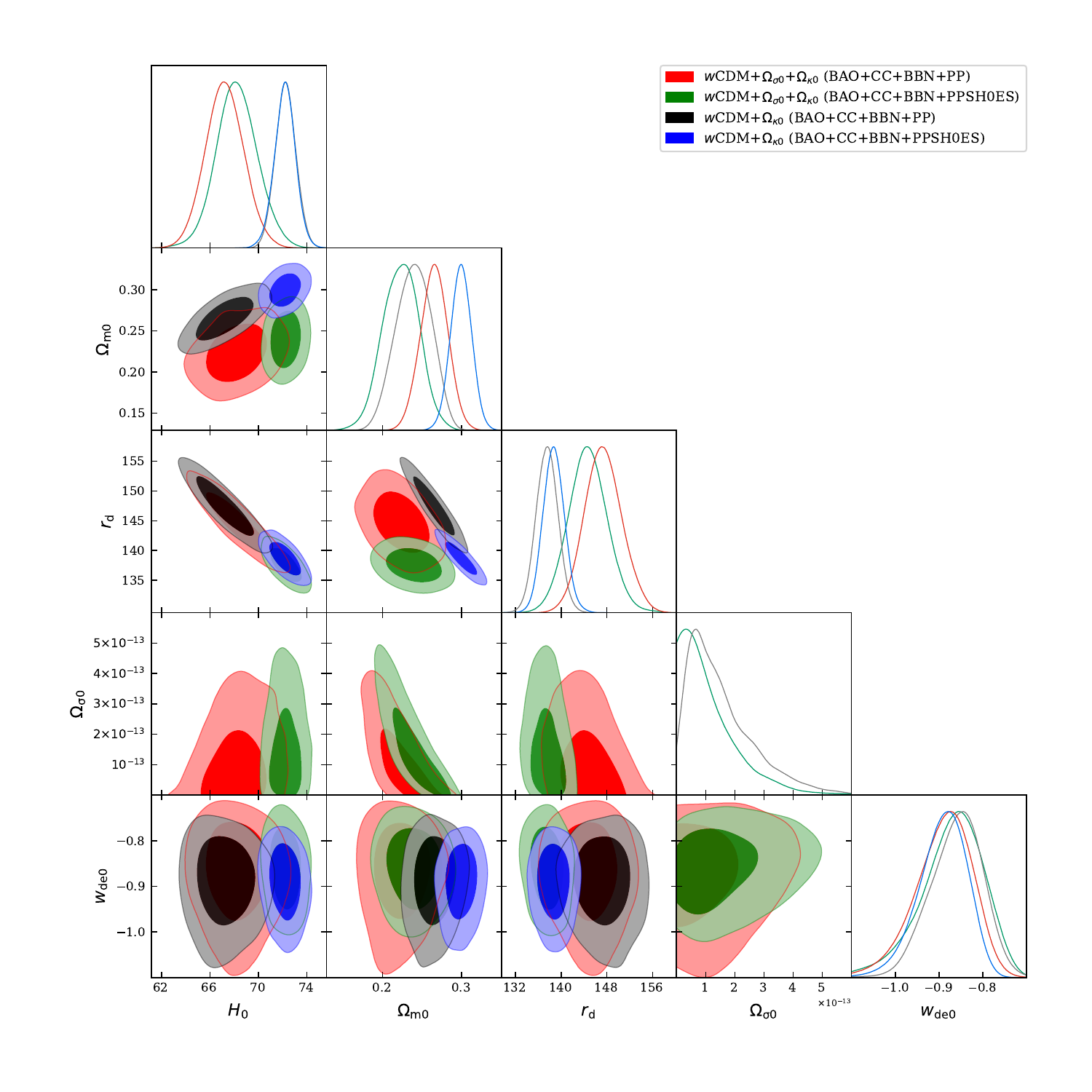}
	\caption{One and $\,\,$ two $\,\,$ dimensional $\,\,$ marginalized $\,\,$ confidence $\,\,$ regions (at $\,\,$ 68\% and $\,\,$ 95\% CL) for some selected parameters  of $\,\,$ $w$CDM+$\Omega_{\rm \kappa0}$ $\,\,$ and $\,\,$ $w$CDM+$\Omega_{\rm \kappa0}$+$\Omega_{\rm \sigma0}$ $\,\,$ models $\,\,$ from $\,\,$ BAO+CC+BBN+PP and $\,\,$ BAO+CC+BBN+PPSH0ES combinations of data.} 
	\label{fig1_5}
\end{figure*}

Consequently, we observe that the $w$CDM+$\Omega_{\rm \kappa0}$+$\Omega_{\rm \sigma0}$ and $\Lambda$CDM+$\Omega_{\rm \kappa0}$+$\Omega_{\rm \sigma0}$ models provide similar upper bounds on anisotropy parameter with both the combinations of data. Also, from left panel of Figure \ref{fig2_5}, we observe a  strong positive correlation between $\Omega_{\rm \kappa0}$ and $\Omega_{\rm \sigma0}$, which indicates that a bigger value of $\Omega_{\rm \kappa0}$ would imply a bigger value of  $\Omega_{\rm \sigma0}$ and vice-versa. For $w$CDM+$\Omega_{\rm \kappa0}$ and $w$CDM+$\Omega_{\rm \kappa0}$+$\Omega_{\rm \sigma0}$ models, the mean values of $w_{\rm de0}$ at 95\% CL lie in the quintessence region ($w_{\rm de0} >-1$) indicating the quintessence behaviour of DE utilizing both combinations of data. The mean value of $w_{\rm de0}$ for $w$CDM+$\Omega_{\rm \kappa0}$ ($w$CDM+$\Omega_{\rm \kappa0}$+$\Omega_{\rm \sigma0}$) model for BAO+CC+BBN+PP combination of data at 68\% and 95\% CL read as $w_{\rm de0} =  -0.892^{+0.076}_{-0.052}$ ($w_{\rm de0} =  -0.871^{+0.084}_{-0.058}$) and $w_{\rm de0} =  -0.89^{+0.13}_{-0.14}$ ($w_{\rm de0} =  -0.87^{+0.14}_{-0.15}$), respectively. No direct correlation of $w_{\rm de0}$ with $H_0$ is found (also evident from Table \ref{tab:1_5}) for both $w$CDM+$\Omega_{\rm \kappa0}$ and $w$CDM+$\Omega_{\rm \kappa0}$+$\Omega_{\rm \sigma0}$ models with both combinations of data. Also, effects of anisotropy can be seen on $w_{\rm de0}$ as $w_{\rm de0}$ finds a positive correlation with $\Omega_{\rm \sigma0}$ (evident from Figure \ref{fig1_5}). In simple words, it  indicates that an elevation in the upper bounds of $\Omega_{\rm \sigma0}$ would lead to a higher value of $w_{\rm de0}$.\\

Now, we focus on $H_0$ values obtained from $w$CDM+$\Omega_{\rm \kappa0}$ and $w$CDM+$\Omega_{\rm \kappa0}$+$\Omega_{\rm \sigma0}$ models utilizing combinations of BAO+CC+BBN+PP and BAO+CC+BBN+PPSH0ES data and do a comparative analysis with the $H_0$ values obtained in \cite{Yadav:2024pvr} for $w$CDM and $w$CDM+$\Omega_{\rm \sigma0}$ models (with the same set of combinations of data). Also, we are not comparing the $H_0$ values obtained from $w$CDM+$\Omega_{\rm \kappa0}$ ($w$CDM+$\Omega_{\rm \kappa0}$+$\Omega_{\rm \sigma0}$) model with $H_0$ values obtained from $\Lambda$CDM+$\Omega_{\rm \kappa0}$ ($\Lambda$CDM+$\Omega_{\rm \kappa0}$+$\Omega_{\rm \sigma0}$) model because their corresponding $H_0$ values with respective combinations of data are almost similar (see Table \ref{tab:1_5}). The constraint on $H_0$ for the $w$CDM+$\Omega_{\rm \kappa0}$ ($w$CDM+$\Omega_{\rm \kappa0}$+$\Omega_{\rm \sigma0}$) model for the BAO+CC+BBN+PP combination of data is as follows: $H_0 = 67.2^{+1.6}_{-1.6} {\rm \,km\,s^{-1}\,Mpc^{-1}} (68.2^{+1.7}_{-1.7} {\rm \,km\,s^{-1}\,Mpc^{-1}})$ at 68\% CL. For same set of combination of data, the constrained value of $H_0$ for $w$CDM ($w$CDM+$\Omega_{\rm \sigma0}$) model \cite{Yadav:2024pvr} is $H_0 = 66.4^{+1.3}_{-1.3} {\rm \,km\,s^{-1}\,Mpc^{-1}} (68.4^{+1.6}_{-1.6} {\rm \,km\,s^{-1}\,Mpc^{-1}})$ at 68\% CL. Similarly, the constrained value of $H_0$ for $w$CDM+$\Omega_{\rm \kappa0}$ ($w$CDM+$\Omega_{\rm \kappa0}$+$\Omega_{\rm \sigma0}$) model with BAO+CC+BBN+PPSH0ES combination of data, is $H_0 = 72.19^{+0.87}_{-0.87} {\rm \,km\,s^{-1}\,Mpc^{-1}}$ $(72.25^{+0.84}_{-0.84} {\rm \,km\,s^{-1}\,Mpc^{-1}})$ at $\,\,$ 68\% CL. For the same combination of data, the constrained value of $H_0$ for $w$CDM ($w$CDM+$\Omega_{\rm \sigma0}$) model \cite{Yadav:2024pvr} reads as $H_0 = 71.51^{+0.83}_{-0.83} {\rm \,km\,s^{-1}\,Mpc^{-1}}$ $(72.24^{+0.82}_{-0.82} {\rm \,km\,s^{-1}\,Mpc^{-1}})$ at 68\% CL. Higher value of $H_0$ is obtained for  BAO+CC+BBN+PP as well as BAO+CC+BBN+PPSH0ES combination of data  when curvature is included in $w$CDM model \cite{Yadav:2024pvr}. Almost similar values of $H_0$ are obtained from $w$CDM+$\Omega_{\rm \sigma0}$ and $w$CDM+$\Omega_{\rm \kappa0}$+$\Omega_{\rm \sigma0}$ models for BAO+CC+BBN+PP as well as BAO+CC+BBN+PPSH0ES combination of data which indicate that anisotropy does not play any important role in case of $w$CDM model when considered with or without curvature. Also, $H_0$ is negatively correlated to $\Omega_{\rm \kappa0}$ or in other words, we can say that a higher value of $H_0$ require a lower value of $\Omega_{\rm \kappa0}$ (as evident from middle panel of Figure \ref{fig2_5}). The highest value of Hubble constant $H_0$ found in this analysis is $72.25 \pm 0.84~{\rm km\, s^{-1}\, Mpc^{-1}}$ (see Table \ref{tab:1_5}) from BAO+CC+BBN+PPSH0ES  combination of data which is found to be consistent with $H_0^{\rm R22}=73.04\pm1.04~{\rm km\, s^{-1}\, Mpc^{-1}}$ from SH0ES measurement. When we quantify the $H_0$ tension with SH0ES measurement for BAO+CC+BBN+PP combination of data, $3.0\sigma$ tension on $H_0$ is found for the $w$CDM+$\Omega_{\rm \kappa0}$ model and $2.4\sigma$ tension on $H_0$ is found for $w$CDM+$\Omega_{\rm \kappa0}$+$\Omega_{\rm \sigma0}$ model. Also, $0.6\sigma$ tension on $H_0$ is seen for $w$CDM+$\Omega_{\rm \kappa0}$ model and $0.5\sigma$ tension is seen for $H_0$ in $w$CDM+$\Omega_{\rm \sigma0}$ model with BAO+CC+BBN+PPSH0ES combination of data. Thus, for BAO+CC+BBN+PP combination of data, $0.6\sigma$ $H_0$ tension is relieved due to the presence of anisotropy of the order $10^{-13}$ with a constant EoS parameter of DE. Also, the inclusion of anisotropy of the same order as above does not contribute much in relieving $H_0$ tension with BAO+CC+BBN+PPSH0ES combination of data. \\

We now go over how $w_{\rm de0}$, $\Omega_{\rm \sigma0}$, and $\Omega_{\rm \kappa0}$ impact $\Omega_{\rm m0}$, the present-day matter density parameter.
For both data combinations, we find that the $w$CDM+$\Omega_{\rm \kappa0}$ model yields considerably lower mean values of $\Omega_{\rm m0}$ than the $\Lambda$CDM+$\Omega_{\rm \kappa0}$ model.
 The mean value of $\Omega_{\rm m0}$ for both data combinations is marginally decreased by introducing anisotropy to the $\Lambda$CDM+$\Omega_{\rm \kappa0}$ model.
 Additionally, adding anisotropy to the $w$CDM+$\Omega_{\rm \kappa0}$ model considerably lowers the $\Omega_{\rm m0}$ mean values for both sets of data.
 For the $w$CDM+$\Omega_{\rm \kappa0}$ model, the constraints on $\Omega_{\rm m0}$ are as follows: $\Omega_{\rm m0 }=0.265\pm 0.017$ and $\Omega_{\rm m0 }= 0.299\pm 0.013$, obtained from the data combinations BAO+CC+BBN+PP and BAO+CC+BBN+PPSH0ES, respectively.
 The constraints on $\Omega_{\rm m0}$ for $w$CDM+$\Omega_{\rm \kappa0}$+$\Omega_{\rm \sigma0}$ model are:  $\Omega_{\rm m0 }=0.224\pm 0.023$ for the combination of data BAO+CC+BBN+PP and $\Omega_{\rm m0 }= 0.239\pm 0.022$ for the combination of data BAO+CC+BBN+PPSH0ES. We also observe that  parameters, $\Omega_{\rm \sigma0}$ and $\Omega_{\rm m0 }$ are negatively correlated (as observed in Figure \ref{fig1_5}) with both combinations of data. \\

Now, we discuss how $\Omega_{\rm m0}$ affects and is affected by other parameters under consideration. We observe a strong negative correlation between $\Omega_{\rm m0}$ and $\Omega_{\rm \kappa0}$ (see right panel of Figure \ref{fig2_5}). It simply means that an increment in the value of curvature parameter would imply a decrement in the value of matter density parameter. Further, the values of $\Omega_{\rm m0}$ for $w$CDM+$\Omega_{\rm \kappa0}$ ($w$CDM+$\Omega_{\rm \kappa0}$+$\Omega_{\rm \sigma0}$) model with BAO+CC+BBN+PP and BAO+CC+BBN+PPSH0ES combinations of data are $\Omega_{\rm m0} =  0.265^{+0.017}_{-0.017}$ ($\Omega_{\rm m0} =  0.224^{+0.023}_{-0.023}$) and $\Omega_{\rm m0} =  0.299^{+0.013}_{-0.013}$ ($\Omega_{\rm m0} =  0.239^{+0.022}_{-0.022}$) respectively. Here, we find that $w$CDM+$\Omega_{\rm \kappa0}$+$\Omega_{\rm \sigma0}$ model provides a lower value of $\Omega_{\rm m0}$ in comparision to the model $w$CDM+$\Omega_{\rm \kappa0}$ with BAO+CC+BBN+PP as  well as BAO+CC+BBN+PPSH0ES combinations of data (see Table \ref{tab:1_5}). Also, we observe a strong negative correlation between $\Omega_{\rm m0}$ and $\Omega_{\rm \sigma0}$ (as evident from Figure \ref{fig1_5}). From here we can deduce that greater the value of $\Omega_{\rm \sigma0}$ parameter, smaller is the value of $\Omega_{\rm m0}$ parameter.

The considered $w$CDM+$\Omega_{\rm \kappa0}$ and $w$CDM+$\Omega_{\rm \kappa0}$+$\Omega_{\rm \sigma0}$ models provide lower values of $r_{\rm d}$ as compared to $\Lambda$CDM+$\Omega_{\rm \kappa0}$ and $\Lambda$CDM+$\Omega_{\rm \kappa0}$+$\Omega_{\rm \sigma0}$ models with BAO+CC+BBN+PP as well as BAO+CC+BBN+PPSH0ES combination of data (the difference in the values of $r_{\rm d}$ from $w$CDM+$\Omega_{\rm \kappa0}$ and $w$CDM+$\Omega_{\rm \kappa0}$+$\Omega_{\rm \sigma0}$ models to $\Lambda$CDM+$\Omega_{\rm \kappa0}$ and $\Lambda$CDM+$\Omega_{\rm \kappa0}$+$\Omega_{\rm \sigma0}$ models is very small in magnitude, which is evident from Table \ref{tab:1_5}). Further, we notice from Table \ref{tab:1_5} that the value of $r_{\rm d}$ reduces while we add anisotropy to $w$CDM+$\Omega_{\rm \kappa0}$ and $\Lambda$CDM+$\Omega_{\rm \kappa0}$ with both data combinations. Also, a negative correlation of $\Omega_{\rm \sigma0}$ to $r_{\rm d}$ is evident from Figure \ref{fig1_5}. We have seen that the inclusion of curvature affects several model parameters. For example, we observe a negative correlation between $H_0$ and $\Omega_{\rm \kappa0}$ (see middle panel of Figure \ref{fig2_5}). Additionally, we see that while fitting the multi-parameter space simultaneously, different parameters influence and are influenced by each other \cite{Marra:2021fvf,Camarena:2023rsd}.
 Curvature specifically affects the other parameters of the examined models (with constant EoS of DE) in a way that relieves the $H_0$ tension up to $\sim 1\sigma$ with BAO+CC+BBN+PP combination of data while no $H_0$ tension is relieved with BAO+CC+BBN+PPSH0ES combination of data.

%%%%%%%%%%%%%%%%%%%%%%%%%%%%%%%%%%%%%%%%%%%%%%%%%%%%%%%%%%%%%%%%%
%%%%%%%%%%%%%%%%%%%%%%%%%%%%%%%%%%%%%%%%%%%%%%%%%%%%%%%%%%%%%%%%%%%%%%%%%    

\section{Conclusion}
\label{sec5}
In this letter, we have reported observational constraints on two $\,\,$ simplest extensions of the standard $\,\,$  $\Lambda$CDM model $\,\,$ with curvature, namely $\,\,$ $w$CDM+$\Omega_{\rm \kappa0}$ $\,\,$ and $\,\,$ $w$CDM+$\Omega_{\rm \kappa0}$+$\Omega_{\rm \sigma0}$ $\,\,$ models from recent low redsift data sets including $\,\,$ BAO, CC, BBN, PP $\,\,$ and $\,\,$ PPSH0ES in two combinations: BAO+CC+BBN+PP and BAO+CC+BBN+PPSH0ES.  We have found that quintessence form of DE is favored for in $w$CDM+$\Omega_{\rm \kappa0}$ as well as $w$CDM+$\Omega_{\rm \kappa0}$+$\Omega_{\rm \sigma0}$ model with both combinations of data at $68\%$ CL as well as $95\%$ CL. Also, a strong positive correlation between $w_{\rm de0}$ and $\Omega_{\rm \sigma0}$ is noted with both  combinations of data, see Figure \ref{fig1_5}. $\Omega_{\rm \sigma0}$ has upper bounds of the order $10^{-13}$ with both the combinations of data. The present day matter density parameter is significantly affected by constant EoS of DE providing lower values for both the models (in comparison to models investigated in \cite{Yadav:2023yyb} with cosmological constant form of DE) with both the combinations of data. Higher values of Hubble constant have been obtained with $H_{\rm 0}=72.19\pm 0.87$ and $H_{\rm 0}=72.25\pm 0.84$ $\rm km\, s^{-1}\, Mpc^{-1}$ at 68\% CL with $w$CDM+$\Omega_{\rm \kappa0}$ and $w$CDM+$\Omega_{\rm \kappa0}$+$\Omega_{\rm \sigma0}$ models, respectively for BAO+CC+BBN+PPSH0ES combination of data. For both the models under consideration with BAO+CC+BBN+PPSH0ES combination of data, the obtained values of $H_0$  are in line with SH0ES measurements ($H_0^{\rm R22}$), and thus relieving the $H_0$ tension. Further, we have observe that adding the anisotropy parameter
 in the $w$CDM+$\Omega_{\rm \kappa0}$ model reduces the $H_0$ tension by $\sim 1\sigma$  for BAO+CC+BBN+PP combination of data. While no effect of anisotropy is observed when included in $w$CDM+$\Omega_{\rm \kappa0}$ model for BAO+CC+BBN+PPSH0ES combination of data. Again, a closed Universe is $\,\,$ favored by $w$CDM as well as $\,\,$ anisotropic $\,\,$ $w$CDM models with $\,\,$ curvature in $\,\,$ analyses with $\,\,$ BAO+CC+BBN+PP combination of data. An observation of an open Universe from $w$CDM model with curvature in analyses with BAO+CC+BBN+PPSH0ES combination of data and a closed Universe from anisotropic $w$CDM model with curvature in analyses with same combination of data is made. The quintessence form of DE is preferred at 95\% CL in both analyses. In summary, our findings demonstrate the impact of anisotropic extensions of the conventional cosmological model in conjunction with curvature on $H_0$ estimation, and also to resolving the $H_0$ tension.
 
\begin{acknowledgments}
 Pardeep gratefully acknowledges the Junior Research Fellowship (File No.~CRG/2021/004658) from the Science and Engineering Research Board (SERB), Govt. of India.  M.Y. is supported by Junior Research Fellowship (File No: 09/1293(15862)/2022-EMR-I) from Council of Scientific and Industrial Research (CSIR), Govt. of India.
 \end{acknowledgments}

\bibliography{refs}

\end{document}